\documentclass[draft,jgrga]{AGUTeX}

\usepackage{graphicx}
\usepackage{lineno}

\usepackage{ulem}
\usepackage{amsmath,xcolor}
\usepackage{indentfirst}
\setkeys{Gin}{draft=false}

\authorrunninghead{Abdullah et al}

\titlerunninghead{Kinetic Ballooning Instability}

\authoraddr{Correspondence to: Ping Zhu\\
Email address: zhup@hust.edu.cn}
\begin{document}


\title{Kinetic Ballooning Instability of the Near-Earth Magnetotail in Voigt Equilibrium}




\authors{Abdullah Khan \altaffilmark{1},
Ping Zhu\altaffilmark{2,3}, Rui Han\altaffilmark{1}, and Ahmad Ali\altaffilmark{1,4}
\altaffilmark{}}
\altaffiltext{1}{CAS Key Laboratory of Geospace Environment and Department of Engineering and Applied Physics, University of Science and Technology of China, Hefei, Anhui 230026, China}
\altaffiltext{2}{International Joint Research Laboratory of Magnetic Confinement Fusion and Plasma Physics, State Key Laboratory of Advanced Electromagnetic Engineering and Technology, School of Electrical and Electronic Engineering, Huazhong University of Science and Technology, Wuhan, Hubei 430074, China}
\altaffiltext{3}{Department of Engineering Physics, University of Wisconsin-Madison, Madison, Wisconsin 53706, USA}
\altaffiltext{4}{National Tokamak Fusion Program, Islamabad, 3329, Pakistan}
 

\begin{abstract}
For a long time, ballooning instabilities have been believed to be a possible triggering mechanism for the onset of substorm and current disruption initiation in the near-Earth magnetotail. Yet the stability of the kinetic ballooning mode (KBM) in a global and realistic magnetotail configuration has not been well examined. In this paper, stability of the KBM is evaluated for the two-dimensional Voigt equilibrium of the near-Earth magnetotail based on an analytical kinetic theory of ballooning instability in the framework of kinetic magnetohydrodynamic (MHD) model, where the kinetic effects such as the finite gyroradius effect, wave-particle resonances, particle drifts motions are  included usually through kinetic closures. The growth rate of the KBM strongly depends on the magnetic field line stiffening factor $S$, which is in turn determined by the effects of the trapped electrons, the finite ion gyroradius, and the magnetic drift motion of charged particles. The KBM is unstable in a finite intermediate range of equatorial $\beta_{eq}$ and only marginally unstable at higher $\beta_{eq}$ regime for higher $T_e/T_i$ values. The finite ion gyroradius and the trapped electron fraction enhance the stiffening factor that tends to stabilize the KBM in the magnetotail far away from Earth. On the other hand, the current sheet thinning destabilizes KBM in the lower $\beta_{eq}$ regime and stabilizes KBM in the higher $\beta_{eq}$ regime.

\end{abstract}


\begin{article}
\section{Introduction}
The ballooning instability of the near-Earth magnetotail has long been suggested as a potential trigger mechanism for the substorm onset. Based on the geostationary satellite (GEOS) 2 observational data, \citet {Roux et al 1991} suggested that the westward traveling surge observed by all-sky cameras was a projection of the ballooning instability during a substorm onset in the equatorial region of the near-Earth magnetotail. Later, other GEOS 2 observational studies further elaborated the role of the ballooning instability as a trigger mechanism for the substorm onset [e.g. Pu et al, 1992, 1997]. Active Magnetospheric Particle Tracer Explorer/Charge Composition Explorer (AMPTE/CCE) satellite observation provides additional evidence that the ballooning instability can be a possible trigger for the substorm onset in the near-Earth magnetotail \citep {Cheng and Lui 1998}. In situ observations suggest that the current sheet breakup in the near-Earth magnetotail can take place prior to the substorm onset in absence of the Earthward fast flow that is often attributed to the middle-magnetotail reconnection \citep {Erickson 2000,Ohtani 2002a,Ohtani 2002b}. \citet {Saito et al 2008} reported the Geotail observations which were consistent with the ballooning mode structures in the equatorial region of the near-Earth magnetotail prior to the substorm onset. Moreover, the observational studies have also demonstrated that the predominant disruption modes prior to the substorm onsets and during the substorm expansion stage possess longitudinal wavelength $\lambda_y$ whose growth rate peaks at $\sim1000 km\sim 0.1 R_E$, corresponding to the wave number $k_y=2\pi/\lambda_y\sim 40 R_E^{-1}$, where $k_y$ is in the direction of dawn-dusk [e.g. \textit{Saito et al, 2008; Liang et al, 2009}]. Recently, \citet{Xing etal 2020} considered ballooning instability as a potential candidate for the cause of auroral wave structures based on THEMIS (Time History of Events and Macroscale Interactions During Substorms) observational data in non-substorm time in the transition region of near-Earth plasma sheet.

Previous analytical and simulation studies regarding the investigation of ballooning instabilities in the near-Earth magnetotail have been mainly based on the ideal MHD model. \citet {Hameiri et al 1991} suggested the existence of ballooning instabilities in the near-Earth magnetotail using the linear MHD theory. The ideal MHD studies indicated that the ballooning instability in the magnetotail would be stable due to strong plasma compression effect \citep {Lee and Wolf 1992,Ohtani and Tamao 1993,Lee and Min 1996}. Both eigenmode and energy principle analyses find stability conditions of different magnetotail configurations for the ideal MHD ballooning mode in the limit of $k_y\rightarrow\infty$ \citep {Lee and Wolf 1992,Pu 1992,Lui et al 1992,Ohtani and Tamao 1993,Dormer 1995,Bhattacharjee 1998,Schindler and Birn 2004,Cheng and Zaharia 2004,Mazur et al 2013}. \citet {Wu et al 1998} and \citet {Zhu 2004} performed linear initial value MHD calculations to study the ballooning modes with finite $k_y$. Their results showed that the growth rate of the ballooning mode increases with $k_y$ and approaches constant value in the large $k_y$ limit. Furthermore, nonideal and kinetic effects were studied for the ballooning instability in the framework of drift MHD, Hall MHD as well as gyrokinetic models \citep{Pu 1997,Cheng and Lui 1998,Lee 1999,Zhu 2003}. In a recent investigation on the pressure anisotropic ballooning instability in the substorm onset mechanism, it is found that the transition from initially perpendicular anisotropy to parallel anisotropy decreases the threshold of plasma beta $(\beta)$ for triggering the ballooning instability \citep{Oberhagemann and Mann 2020}.

Besides the ballooning instability, magnetic reconnection is another candidate for the substorm onset trigger mechanism. \citet{Hones 1977} proposed a substorm onset scenario in the near-Earth magnetotail region that the neutral $X$ line and the resulting plasmoid formation could be developed close to the end of the substorm growth phase, probably due to the onset of instability in tearing mode. For the configuration of the Earth's magnetotail, the formation of plasmoids was also studied in other similar scenarios of substorm onset \citep{Birn and Hones 1981,Lee at al 1985,Hautz and Scholer 1987,Otto et al 1990,Zhu and Raeder 2014}. Observations indicate the presence of $B_z$ minimum width during the substorm growth phase of the near-Earth magnetotail \citep{Sergeev et al 1994,Saito et al 2010}, which may be the result of adiabatic magnetotail convection \citep{Erickson and Wolf 1980}, or the magnetic flux transport caused by the dayside magnetic reconnection \citep{Otto and Hsieh 2012}, where $B_z$ is the magnetic field normal to the neutral sheet or the equatorial plane at $z = 0$. Such a minimum $B_z$ configuration of the near-Earth magnetotail has been found crucial for the onsets of both ballooning instability and plasmoid formation \citep{Zhu and Raeder 2013,Zhu and Raeder 2014}.

\citet{Cheng and Lui 1998}, \citet{Cheng and Gorelenkov 2004}, and \citet{Cheng 2004} considered low-frequency electromagnetic perturbations with $\omega\ll\omega_{ci}$, $k_{\perp}\gg k_{\parallel}$, $k_{\perp}\rho_i\sim{1}$, $v_e>(\omega/k_{\parallel})>v_i$ where $\omega$ is the wave frequency, $\omega_{ci}$ is the ion cyclotron frequency,  $k_{\perp}(k_{\parallel})$ are the perpendicular (parallel) wavenumbers, $\rho_i$ is the ion gyroradius, and $v_e(v_i)$ are the electron (ion) thermal speeds. They further adopted the local approximation for the KBM analysis to be such that the gyroradius effect for electrons is ignored and to be assumed that $|v_{\parallel}\nabla_{\parallel}|\gg\omega,\omega_{de}$, where $\omega_{de}$ is the electron diamagnetic drift frequency. The untrapped electron dynamic is determined by its fast parallel transit motion along the magnetic field $\boldsymbol{B}$, and to the lowest order in $\omega/|v_{\parallel}\nabla_{\parallel}|$, while the trapped electron dynamic is determined by its fast parallel bounce motion and to the lowest order in $\omega/\omega_{be}$, where $\omega_{be}$ is the bounce electron frequency. Furthermore, the untrapped electron motion perpendicular to $\boldsymbol{B}$ is signified mainly by the drift motion of $\boldsymbol{E}\times\boldsymbol{B}$ while the trapped electron motion perpendicular to $\boldsymbol{B}$ is determined by the $\boldsymbol{E}\times\boldsymbol{B}$ drift motion and the magnetic drift motion. Note that the ion dynamic is significantly different from electron perpendicular motion if $k_y\rho_i\sim 1$. The ion dynamic is mainly determined by the polarization drift, $\boldsymbol{E}\times\boldsymbol{B}$ drift and magnetic drift motions, where $\omega,\omega_{di}\gg|v_{\parallel}\nabla_{\parallel}|$, and $\omega_{di}$ is the ion diamagnetic drift frequency.

The kinetic MHD model has been developed in several previous studies of the ballooning instability \citep {Cheng and Lui 1998,Wong et al 2001,Horton et al 1999,Horton et al 2001}. Kinetic effects such as trapped particle dynamics, FLR, wave-particle resonance and parallel electric field were shown to be important in determining the stability of KBM in the magnetosphere \citep {Cheng and Lui 1998}. Their theory is able to explicate the wave frequency, growth rate, and the high critical $\beta$ threshold of low-frequency global instability observed by AMPTE/CCE, where $\beta$ is the ratio of thermal pressure to the magnetic pressure. Low frequency instability is the instability in which the wave frequency is smaller than the particle cyclotron frequency but greater than bounce frequency, i.e. $\omega_{b}\ll\omega\ll\omega_{c}$, where $\omega_{c}$ and $\omega_{b}$ are the particle gyrofrequency and bounce frequency, respectively. In particular, the trapped electron effect coupled with FLR effect produces a large parallel electric field that enhances parallel current and results in much higher stabilizing field line tension than predicted by the ideal MHD theory. As a result, a much higher critical value of $\beta$ than that predicted by the ideal MHD model was obtained. Furthermore, \citet {Horton et al 2001} investigated the stability of ballooning mode in geotail plasma and compared the calculation results between kinetic stability and MHD stability. \citet {Wong et al 2001} developed a general stability theory of kinetic MHD for drift modes at both low and high $\beta$ limits. These kinetic MHD studies provide a better understanding of ballooning instability in wider parameter regimes of the near-Earth magnetotail plasmas. However, the stability of the KBM in a global and realistic configuration has been less studied.

This study is thus devoted to the evaluation of the KBM stability of a global near-Earth magnetotail configuration using the more realistic 2D Voigt equilibrium model. Previously, \citet {Zhu 2003,Zhu 2004, Zhu 2007} and \citet {Lee 1998,Lee 1999} have used the Voigt equilibrium model for investigating the ballooning mode stability of magnetotail configuration within the framework of ideal, Hall and extended MHD models. In this paper, we study the stability of KBM in the global magnetotail plasmas based on the Voigt equilibrium model for the first time. The mechanism responsible for the KBM stabilization in collisionless magnetotail plasmas is found to be the combined effects of trapped electron dynamics, the finite ion gyroradius (FLR), and diamagnetic drifts. Analysis shows that the KBM growth rate strongly depends on the parameter $S(=1+(n_e/n_{eu})\delta)$; where $n_e$ and $n_{eu}$ are the total and untrapped electron densities, respectively. The parameter $\delta$ in the stiffening factor $S$ is defined in terms of the finite ion Larmor radius, electron $\beta_e$, and diamagnetic drifts, as in equation (5) of Section 3. The stiffening factor $S$ depends on the untrapped/trapped electron dynamics, FLR effect and magnetic drift motion of the charged particles. Stability of the KBM is evaluated in a broad range of $\beta_{eq}$ at different $k_y$, $T_e/T_i$, and current sheet width, where $\beta_{eq}$ represent the value of $\beta$ at equatorial plane. Our results suggest that the excitation of KBM instability through current sheet thinning remains a viable scenario for substorm onset in the near-Earth magnetotail.

The rest of the paper is organized as follows. The Voigt equilibrium model for magnetotail plasma is presented in Section 2. In Section 3, we briefly recount the dispersion relation of the ballooning mode based on an analytical kinetic theory. In Section 4, we evaluate the growth rate of kinetic ballooning mode from theory for the Voigt equilibrium. Finally, a summary of the key results and a brief discussion on the new aspects of this study are presented in Section 5.

\section{Voigt Model of Magnetotail Equilibrium}
In the 2D Voigt equilibrium model, the magnetic field lies in x-z plane can be expressed in terms of magnetic flux function $\Psi$ as $\bf B$ $=-\boldsymbol\nabla\times(\Psi \hat y)=\hat y \times\boldsymbol\nabla\Psi$ \citep {Voigt 1986}. Previously, this model has been applied in the ideal and Hall MHD studies of the ballooning instabilities in the magnetotail plasmas \citep {Zhu 2003,Zhu 2004}. Since there is no X-point in the magnetic configuration, therefore, the role of pre-existing magnetic reconnection process may be eliminated. Moreover in the Voigt model, the equilibrium current density parallel to the magnetic field lines is kept zero to exclude the current driven instabilities. This is achieved in the Voigt equilibrium model by setting the magnetic field scaling parameter $h=0$. Therefore, in the Voigt equilibrium model, the only driving force for the ballooning instability in the bad curvature region of the magnetic field lines comes from the pressure gradient. Finally, the equilibrium current sheet thickness in the Voigt model could be varied by adjusting the equilibrium model parameters which allows the equilibrium magnetic field lines to vary from dipole-like to tail-like for a long range of plasma $\beta$. 

In this study, the geocentric solar magnetospheric (GSM) coordinates (x, y, z) are used, where $x$ is directed towards the sun, $y$ is from dawn-to-dusk, and $z$ is in the northward direction in the equatorial plane defined by the magnetic dipole axis of the Earth. For ballooning modes with large $k_y$, the near-Earth region of the magnetotail can be modeled using the 2D magnetostatic Voigt equilibrium \citep {Voigt 1986,Hilmer and Voigt 1987,Voigt and Wolf 1988}. In this model, the Grad-Shafranov equation includes the 2D dipole magnetic field of the Earth and can be expressed as follows:
\begin{eqnarray}
\nabla^2 \Psi + \frac{d}{d\Psi}[\mu_0P(\Psi) +\frac{1}{2}B_y^2(\Psi)]=-M_D\frac{\partial}{\partial x}\delta(x)\delta(z).
\end{eqnarray}
If we consider the Voigt equilibria $P(\Psi)=k^2\Psi^2/2\mu_0$ and $B_y(\Psi)=-h\Psi$ then equation (1) can be solved analytically, where $P$ is the equilibrium pressure and $B_y$ is the equilibrium magnetic field in the direction of dusk-dawn, $k$ and $h$ are the constant scaling parameters. Note that, in this study we assume $B_y=0$ by setting $h=0$. The Grad-Shafranov equation is linear in $\Psi$ with Voigt equilibria and its solution for the night-side magnetosphere $(x<0)$ takes the following form
\begin{eqnarray}
\Psi(x,z)=-\frac{M_DR_E}{2}\sum_{n=1}^{\infty}\cos(\eta_n z) e^{\lambda_n x} (1+e^{-2\lambda_n x_b})+\Psi_{-\infty}
\end{eqnarray}
where $M_D$ is the dipole moment of the Earth and $R_E$ is the Earth's radius. The eigenvalues $\lambda_n$ and $\eta_n$ are related by two free physical parameters $h$ and $k$ as $\lambda_n^2=\eta_n^2-k^2-h^2$, where $\eta_n=({\pi}/{2})({2n-1})/z_{mp}$. The tail and day-side magnetopause locations are $z_{mp}$ and $x_b$ respectively. The magnetic flux function ($\Psi_{-\infty}$) at $x\rightarrow{-\infty}$ is considered zero in this study. The key parameters in the Voigt equilibrium model are $x_b$, $z_{mp}$, $k^2$ and $h$. The magnetic field line configuration of the 2D Voigt equilibrium is shown in figure 1. We choose two different values of the pressure scaling parameter $k^2 (0.15$ and $0.272)$ to demonstrate the actual stretching of magnetic field lines in the near-Earth magnetotail. For low values of $k^2$, the magnetic field lines are round-shaped i.e. dipole-like (figure 1a). As $k^2$ increases the field lines become stretched i.e. tail-like (figure 1b).
\section{Dispersion Relation of KBM}
The kinetic ballooning instability perturbations are considered in the regime $k_\perp \rho_i=O(1)$ and $k_\parallel\ll k_\perp$, where $k_\parallel$ and $k_\perp$ represent the parallel and perpendicular wave number, respectively. A local dispersion relation for the kinetic ballooning instability was derived for the frequency ordering $\omega\gg\omega_{de},\omega_{di}$. Neglecting the nonadiabatic density and pressure responses, the dispersion relation is obtained \citep {Cheng and Lui 1998}
\begin{eqnarray}
\omega(\omega-\omega_{\star pi})\simeq ({S}{k_{\parallel}^2}-\frac{\beta_{eq}}{L_pR_c}){(1+b_i)V_A^2},
\end{eqnarray}
Here $S$ is defined as follows \citep {Cheng and Gorelenkov 2004}:
\begin{eqnarray}
S=1+\frac{n_e}{n_{eu}}\delta,
\end{eqnarray}
\begin{eqnarray}
\delta&=& \frac{\beta_e}{2}(\frac{\omega_{\star pi}-\omega_{\star pe}}{\omega})^2-\frac{q_i T_e}{q_e T_i}(\frac{\omega-\omega_{\star pi}}{\omega-\omega_{\star e}})b_i-\frac{3}{2}(\frac{\omega-\omega_{\star pe}}{\omega-\omega_{\star e}})\frac{<\hat\omega_{de}>}{\omega} \\ \nonumber
&+&(\frac{\omega-\omega_{\star pi}}{\omega-\omega_{\star e}})\frac{\hat\omega_{Be}+\hat\omega_{Ke}}{2\omega},
\end{eqnarray}\\
where $R_c$ and $L_p$ are the radius of the magnetic field curvature and the pressure gradient scale length i.e. $\beta_{eq}/L_pR_c=2\mu_0\boldsymbol\kappa\cdot\boldsymbol\nabla P/B^2$, where $\boldsymbol\kappa=\hat{b}\cdot\boldsymbol\nabla\hat{b}$ is the magnetic field curvature with $\hat{b}=\boldsymbol{B}/B$, $V_A=B/\sqrt{\mu_0n_i m_i}$ is the Alfv\'en velocity, $b_i=k_y^2\rho_i^2/2$, $\rho_i=v_{thi}/\omega_{ci}$, $v_{thi}=(2T_i/m_i)^{1/2}$, $\omega_{ci}=eB/m_i$, $T_i=P/n(1+T_e/T_i)$, $n_{eu}/n_e=1-(1-B(x_e)/B_{max})^{1/2}$ is the ratio of untrapped electron density to total electron density, $B(x_e)$ is the magnetic field at the equatorial location $(x=x_e, z=0)$, $B_{max}$ denotes the magnetic field amplitude at the effective ionosphere-magnetosphere boundary defined as a sphere with the radius $r_{bc}$ (Figure 1), and $n_e/n_{eu}\simeq 2B_{max}/B(x_e)\gg{1}$ near the equator. To evaluate the KBM growth rate, we numerically solve the two nonlinear equations (3) and (4). Since the wave frequency $\omega(=\omega_r+i\gamma_k)$ is a complex number, therefore the stiffening factor $S$ is complex number as well, where $\omega_r$ is the real frequency of the KBM and $\gamma_k$ is  the KBM growth rate. In equation (4) $\beta_e=2\mu_0n_eT_e/B^2$ represents the plasma parameter $\beta$ for electron and $\omega_{\star pj}=\omega_{\star j}(1+\eta_j)$ is the diamagnetic drift frequency, where $\omega_{\star j}={\bf B}\times\boldsymbol\nabla P_j\cdot{\bf k_{\perp}}/(B\omega_{cj} P_j)({T_j}/{m_j})$ and $\eta_j=d\ln T_j/d\ln n_j$. Furthermore, $\hat\omega_{Be}=2{\bf B}\times\boldsymbol\nabla B\cdot{\bf k_{\perp}}T_e/q_e B^3$ and $\hat\omega_{Ke}=2{\bf B}\times{\boldsymbol\kappa}\cdot{\bf k_{\perp}}T_e/q_e B^2$. The third term in the square bracket of equation (4) is neglected in these calculations as the wave frequency $\omega$ is much larger than the bounce-average of $\omega_{de}$ i.e. $(<\omega_{de}>)$ \citep{Cheng and Lui 1998}.\\

\section{Key Parameter Dependences of KBM Growth Rate}
In this section, we evaluate the dependence of KBM growth rate on the key parameters in the near-Earth magnetotail, including the local $\beta_{eq}$, the ion Larmor radius, and the trapped electron fraction, among others. The key parameters for the KBM dispersion relation at the equatorial plane are evaluated as a function of $x_e$ from $x_e=-5 R_E$ to $x_e=-16 R_E$ at $k^2=0.15$ and $k^2=0.272$ for the Voigt equilibrium model. Figure 2 shows that $L_p=P/(dP/dx)$, $\beta_{eq}$, the finite ion Larmor radius $(\rho_i)$ all rise with $k^2$ from the lowest values at $x_e=-5 R_E$ to the highest values as soon as $x_e\leq -9 R_E$. Similarly, $\beta^{MHD}=k_{\parallel}^2L_pR_c$ increases with $k^2$ from the higher value at $x_e=-5 R_E$ to the lowest value once $x_e\leq -9 R_E$. The $R_c^{-1}$ varies very slowly (straight lines), whereas $k_{\parallel}$ decreases rapidly along the x-axis tailward, and $k_{\parallel}$ becomes less than $R_c^{-1}$ at equatorial locations $(x_e)$ farther away from Earth (Figure 2d). $\beta^{MHD}$ is the $\beta$ threshold, above which the ballooning mode become unsatble within the ideal incompressible MHD model, i.e. $k_{\parallel}^2=\beta^{MHD}/L_pR_c$. Moreover, $\beta^{MHD}$ is obtained from $k_{\parallel}^2=\beta^{MHD}/L_pR_c$ when setting equation 3 to zero within the ideal MHD model. It is the critical $\beta$ value, above which the line bending force represented by $k_{\parallel}^2$ is overcome by the interchange force represented by $\beta/(L_p R_c)$. The parallel wave number measures from the field line configuration by $2\pi/L(x_e)$ with $L(x_e)$ being two times the length of field line starting from the crossing point $(x_e,0)$ at the equatorial plane to the effective ionosphere-magnetosphere boundary location represented by the $r_{bc}$ circle originated at $(0,0)$ in the x-z plane, with the radius $r_{bc}=4.65 R_E$ (Figure 1). This choice of the effective $r_{bc}$ or $k_{\parallel}$ is mainly based on the observations and previous MHD analyses that the near-Earth magnetotail region near the geosynchronous orbit (i.e. $6.6 R_E$) is stable to macroscopic MHD type of perturbations, including the ballooning modes in the spectrum regime where the parallel wavelength is greater than $L(x_e)$ for $x_e\leq 6.6 R_E$ (see also Figure 8). Indeed, the equatorial beta $\beta_{eq}$ is lower at $x_e=-5 R_E$, continues to increase with $x_e$, and becomes leveled out beyond $x_e=-9 R_E$, whereas critical value $\beta^{MHD}$ tends to vary inversely along the x-axis (Figure 2b). The trapped electron fraction $n_{et}/n_e(=1-B/B_{max})^{1/2}$ increases with $x_e$ moving away from the Earth for both thinner and wider current sheets. The trapped electrons are the electrons that are bounced back at any location of the field lines, whereas those who can pass through the location are considered as passing electrons.

For all calculation results, the ion and electron number density are taken as $n=1cm^{-3}$ in the central plasma sheet. The dipole moment of the Earth $M_D$ is set to be $4000 nT$ based on the expected value of the Earth's dipole field and the realistic variations of magnetic field and plasma pressure at the equatorial plane. The temperature ratio $(T_e/T_i)$ in equation (4) is assumed to be less than unity based on observations. Different values of $T_e/T_i(=0.1-0.4)$ are considered to investigate its effect on the KBM growth rate.
\subsection{Dependence on Local Parameters}
\textbf{\textit{--Stiffening Factor $\boldsymbol{(S)}$}}

To isolate the dependence of KBM growth rate on the stiffening factor $S$ at the equatorial points $x_e=-9 R_E$ and $z=0$ with $T_e/T_i=0.1$ and $k_y=40 R_E^{-1}$, we solve equation (3) for the KBM growth rate with $S$ fixed as a parameter. The KBM growth rate is usually at its maximum at $k_y=40 R_E^{-1}$ which corresponds to the wave length $\lambda_y\sim 0.1R_E$ that is on the order of typical ion gyroradius (Figure 6). This is also supported by recent observations on the auroral bead structure, which maps to a wave-like structure in magnetotail with a wavelength $\lambda_y\sim 600-1000km$, corresponding to the wave number $k_y=2\pi/\lambda_y$ in the regime of $\sim 40-67 (R_E^{-1})$ [e.g. \textit{Saito et al}, 2008; \textit{Xing et al}., 2020]. As $S$ increases, the KBM growth rate quickly reduces to zero when $S\sim{160}$ for both $k^2=0.15$ and $k^2=0.272$ (Figure 3), confirming the strong stabilizing effect from the stiffening factor $S$  \citep{Cheng and Lui 1998}. However, the factor $S$ is actually a function of several other more fundamental parameters such as the finite ion Larmor radius and the diamagnetic drift velocity, which may have both direct and indirect effects on KBM either through or not through $S$.  We look into these other key parameters next.

\textbf{\textit{--Finite ion Larmor Radius $(\boldsymbol{\rho_i})$}}

The finite ion Larmor radius $\rho_i$ appears in $b_i=k_y^2\rho_i^2/2$ in both equations (3) and (4), which are solved together for the real frequency and growth rate of KBM. For the wider current sheet $(k^2=0.15)$ with $T_e/T_i=0.1$ and $k_y=40 R_E^{-1}$ at the equatorial points $x_e=-9 R_E$ and $z=0$, the KBM growth rate increases with $b_i$ when $b_i\le{0.3}$, and eventually becomes fully suppressed when $b_i$ is above 2.78 (Figure 4-a). Such a FLR stabilization of KBM has to act through the stiffening factor $S$, which would be totally absent if $S$ is artificially kept fixed (Figure 4). For the thinner current sheet with $k^2=0.272$, the dependence of KBM growth rate on $\rho_i$ is similar (Figure 4-b).

\textbf{\textit{--Trapped Electron Fraction $\boldsymbol{(n_{et}/n_e)}$}}

Equations (3) and (4) are also solved simultaneously to evaluate the KBM growth rate as a function of trapped electron fraction $(n_{et}/n_e)$ at $k_y=40 R_E^{-1}$ and $x_e=-9R_E$ for two different typical values of the pressure scaling parameter $(k^2)$ (Figure 5). Here the trapped electron fraction $n_{et}/n_e$ is varied as a parameter instead of a function of $x_e$. For the wider current sheet $(k^2=0.15)$, there is only a weak dependence of KBM growth rate on $n_{et}/n_e$; however, for the thinner current sheet $(k^2=0.272)$, the effect of $n_{et}/n_e$ on KBM growth rate becomes more apparent. The KBM growth rate against $n_{et}/n_e$ for the thinner current sheet becomes zero when $n_{et}/n_e\sim 9.5$. In both cases, the trapped electron fraction itself tends to stabilize KBM.

\textbf{\textit{--$\boldsymbol{k_y}$ Dependence (Diamagnetic Effects)}}

We evaluate the growth rate of the KBM in a long range of perpendicular wave number $(k_y)$ normalized with an ion gyroradius $(\rho_i)$ for different values of the pressure scaling parameter $k^2$ for temperature ratio $T_e/T_i=0.1$ at the equatorial location $x_e=-9 R_E$ and $z=0$. The KBM is found to be unstable at low end of $k_y\rho_i$ for various choices of $k^2$ where the ideal MHD effects are dominant over the kinetic effects. For $k^2=0.15$ case, the KBM growth rate is almost constant until becomes suppressed when $k_y\rho_i>2.9$. For higher values of $k^2$ which corresponds to stretched thin current sheet configurations, the KBM growth rate is more prominent for all range of $k_y\rho_i$ (figure 6).\\

\noindent\textbf{\textit{--$\boldsymbol{\beta_{eq}}$ Dependence}}

Figure 2 shows that the $\beta_{eq}$ variation on equatorial plane is rather small beyond $9 R_E$, thus in figure 7, the $\beta_{eq}$ value at $x_e=-9 R_E$ is used as a representative of the overall $\beta$ level of equilibrium. For the KBM analysis in the near-Earth magnetotail, the pressure scaling parameter $(k^2)$ is used to control the local $\beta_{eq}$. The plasma $\beta_{eq}$ increases with $k^2$ and the magnetic field lines become more stretched and tail-like. The KBM growth rate as a function of $\beta_{eq}$ through the equatorial point $x_e=-9 R_E$ with $\eta_j=0$ is plotted for different values of $k_y$ shown in figure 7. Consistent with the previous subsection, the KBM growth rate is generally lower at higher values of $k_y$ for a given $\beta_{eq}$. The KBM growth rate increases with $\beta_{eq}$ in the lower $\beta_{eq}$ regime where the pressure-driven term $\beta_{eq}/L_pR_c\sim 2\mu_0\boldsymbol\kappa\cdot\boldsymbol\nabla P/B^2$ is dominant. Since the FLR $(\rho_i)$ increases with the pressure scaling parameter $k^2$ as $\rho_i\propto\beta_{eq}^{1/2}$ (shown in Figure 2-c), which in turn increases the stiffening factor $S$, the KBM growth rate decreases in the higher $\beta_{eq}$ regime due to the FLR stabilization through $S$. Figure 7-b shows the variation of the KBM growth rate over a broad spectrum of $\beta_{eq}$ for different choices of $T_e/T_i$ at $k_y=38 R_E^{-1}$, indicating non-monotonic effects of the $T_e/T_i$ ratio. 
\subsection{Dependence on equatorial location and current sheet width}

We evaluate the maximum growth rate of the KBM as a function of $x_e$ for two typical values of pressure scaling parameter $k^2=0.15$ and $k^2=0.272$ at $T_e/T_i=0.1$ and $k_y=38R_E^{-1}$ (Figure 8). The profiles of number density $n(x_e)$ for both cases are shown in figure 8-a. For $k^2=0.15$ (wider current sheet case), the number density profile is obtained using $n=P/(T_e+T_i)$, there the typical value of $T_i$ in the near-Earth plasma sheet is taken as 1 keV \citep{Kivelson and Russell 1995}. The value of ion temperature at $1keV$ corresponds to the realistic number density profile (see Figure 8-a) in a low $\beta_{eq}$ regime, which is not the only or the representative regime for the near-Earth plasma sheet at all times. It is considered here in the study to compare with the more common higher $\beta_{eq}$ regimes, where $T_i$ is typically higher than $1keV$ (see Figure 7). For $k^2=0.272$ (thin current sheet case), the number density profile is kept same with the wider current sheet case, while the corresponding ion temperature is obtained from $T_i=P/n(1+T_e/T_i)$. The stiffening factor increases significantly with the equatorial location $x_e$ (Figure 8-b). For the wider current sheet, the KBM growth rate increases in the region $5<-x_e<10$ due to the dominant effect of the ballooning drive term $-\beta_{eq}/L_pR_c$ (Figure 8-c). The KBM growth rate reaches to the peak value at $x_e\sim-10.2 R_E$  and then decreases with the stiffening factor $S$ tailward. For the thin current sheet configuration $(k^2=0.272)$, the KBM growth rate varies along $x_e$ similarly, however, the magnitude of which is nearly 200 times larger than that in the wider current sheet case (Figure 8-c).

The parameter $z_{mp}$ defines the magnetopause location along z-direction and we use the parameter $z_{mp}$ as a proxy for the current sheet thickness in the Voigt model of magnetotail equilibrium. For $k_y=38 (R_E^{-1})$ and $T_e/T_i=0.1$, we compare the $\beta_{eq}$ dependence of KBM growth rate between the thin $(z_{mp}=1 R_E)$ and the wider current sheets $(z_{mp}=3 R_E)$ (Figure 9). The KBM in the thin current sheet configuration is significantly more unstable, suggesting that the two possible scenarios proposed in ideal MHD model for the substorm onset trigger through ballooning instability remains possible even in the regime of KBM \citep {Zhu 2004}.
\section {Summary and Discussion}
In this paper, a local dispersion relation for the KBM stability is evaluated for the near-Earth magnetotail in a broad range of key plasma sheet parameters including the finite ion Larmor radius $(\rho_i)$ and equatorial beta $(\beta_{eq})$ for both dipole-like and stretched 2D Voigt equilibriums, which are meant for modeling the near-Earth magnetotail configuration during the slow substorm growth phase. Our results show that the growth rate of KBM is strongly dependent on the magnetic field stiffening factor $S$, which is mainly determined by the trapped/untrapped electron fraction, the finite ion gyroradius, and the magnetic drift motion of charged particles. It is found that KBM is most unstable in the intermediate equatorial $\beta_{eq}$ for $k_y\rho_i<1$ and in the tail region $(10-11) R_E$ due to the combined stabilizing effects from the finite ion Larmor radius and the trapped electrons elsewhere. Our results also indicate that the current sheet thinning enhances the KBM growth rate and such a mechanism remains a highly trigger for the substorm onset in the near-Earth magnetotail.

The KBM stability of the near-Earth magnetotail in other types of configurations prior to the substorm onset needs to be examined in future work. Furthermore, \citet{Cheng and Lui 1998} used local approximations to develop the kinetic ballooning instability theory. In future study, the global eigenmode analysis approach is also worth exploring.


\acknowledgments
This research was supported by the Fundamental Research Funds for the Central Universities at Huazhong University of Science and Technology Grant No.2019kfyXJJS193, the National Natural Science Foundation of China Grant Nos. 41474143 and 51821005. The author Abdullah Khan acknowledges University of Science and Technology of China for awarding the Chinese Government Scholarship for his Ph.D. study. The author A. Ali acknowledges the support of the State Administration of Foreign Experts Affairs--Foreign Talented Youth Introduction Plan under Grant No. WQ2017ZGKX065. This work has not used any previous or new data.

\clearpage
\newpage
\begin{figure}
\hspace*{-1.5cm}
\centering
\includegraphics[height=17 cm,width=19 cm]{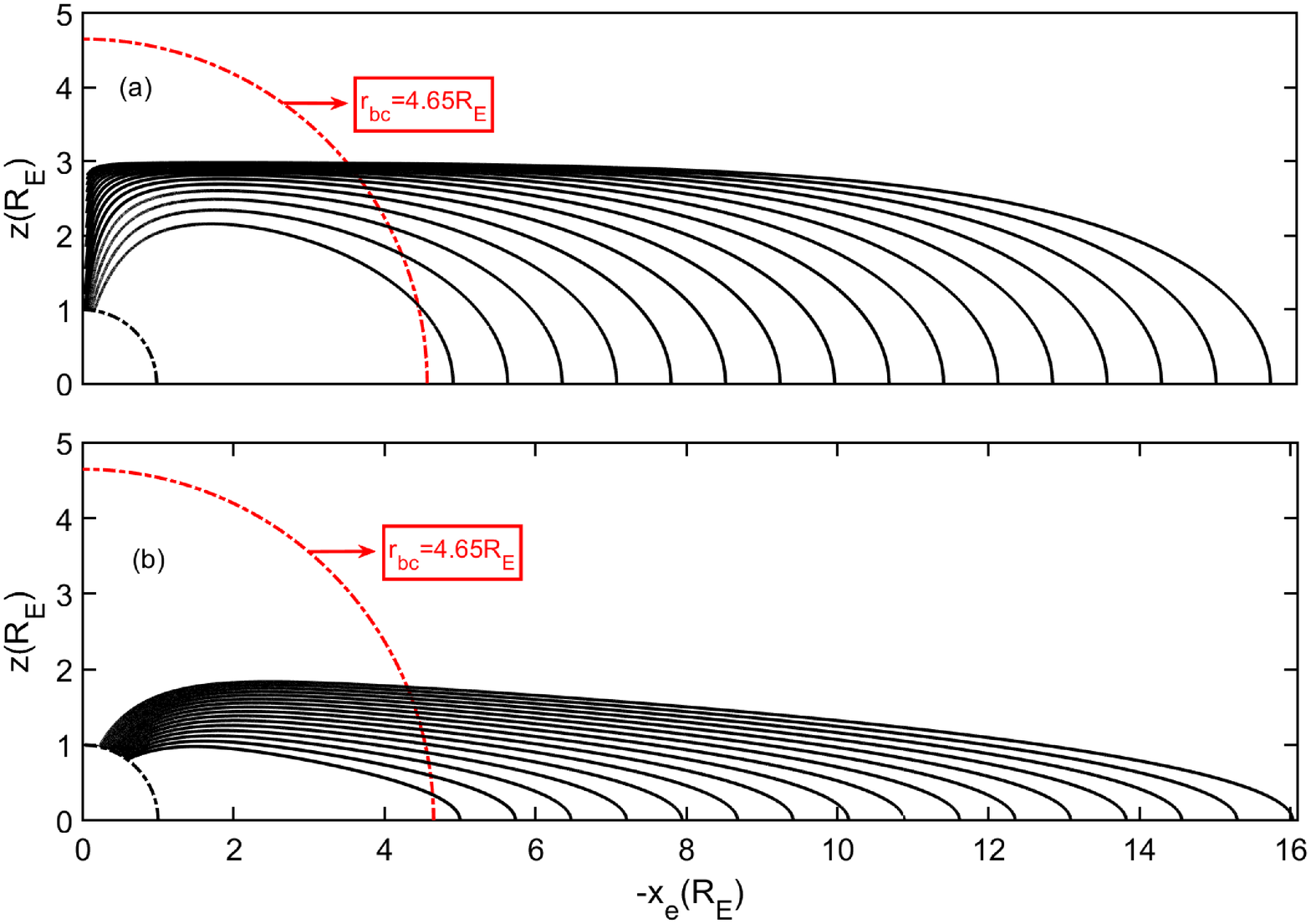}
 \caption{\label{pre_spit} Magnetic field lines of the Voigt equilibrium in the near-Earth magnetotail. The equilibrium parameters are $x_b=6, z_{mp}=3$ (a) $k^2=$0.15, and (b) $k^2=$0.272.}
\end{figure}
\clearpage
\begin{figure}[htbp]
\hspace*{-1.6cm}
\centering
\includegraphics[height=17.5cm,width=20 cm]{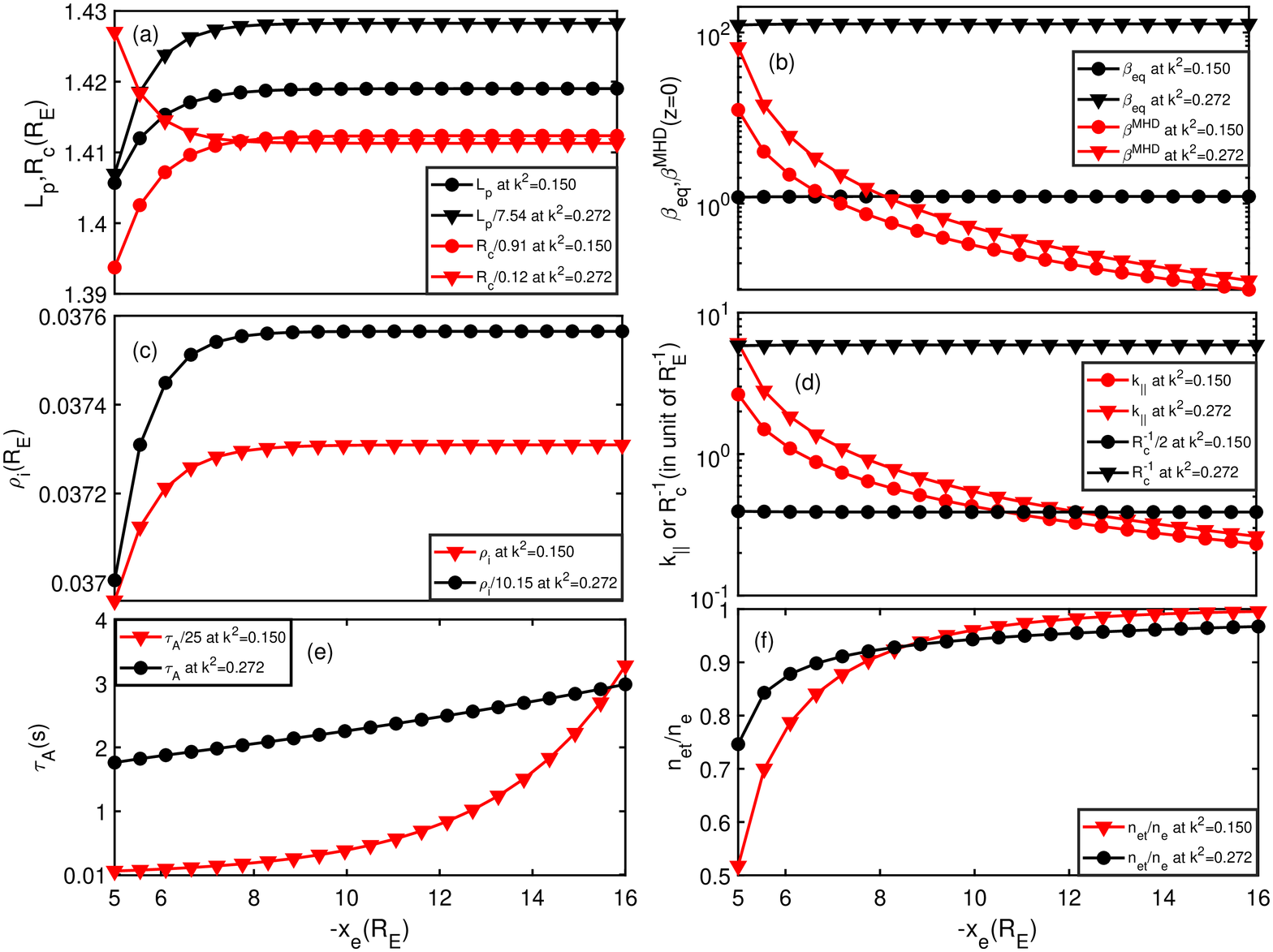}
\caption{\label{pre_spit}Variation of the key stability parameters as a function of $x_e$ for two different $k^2$ values at $z=0$ using the Voigt equilibrium model (a) the plasma scale length $L_p=P/(dP/dx)$ and the radius of curvature $R_c=1/|\hat{b}\cdot \nabla \hat{b}|$, (b) the plasma parameter $\beta_{eq}=2P/B^2$ and the $\beta^{MHD}=k_{\parallel}^2L_pR_c$ (the critical $\beta$ value for the onset of incompressible ballooning instability in the ideal MHD model), (c) the finite ion gyroradius $\rho_i$, (d) the magnetic field curvature $R_c^{-1}=|\hat{b}\cdot\nabla\hat{b}|$ and the parallel wave number $k_{\parallel}=2\pi/L(x_e)$, (e) the Alfv\'en time $\tau_A=R_E/V_A$, and (f) the trapped electron fraction $n_{et}/n_e$.}
\end{figure}
\clearpage
\begin{figure}
\hspace*{-1.5cm}
\centering
\includegraphics[height=15 cm,width=19 cm]{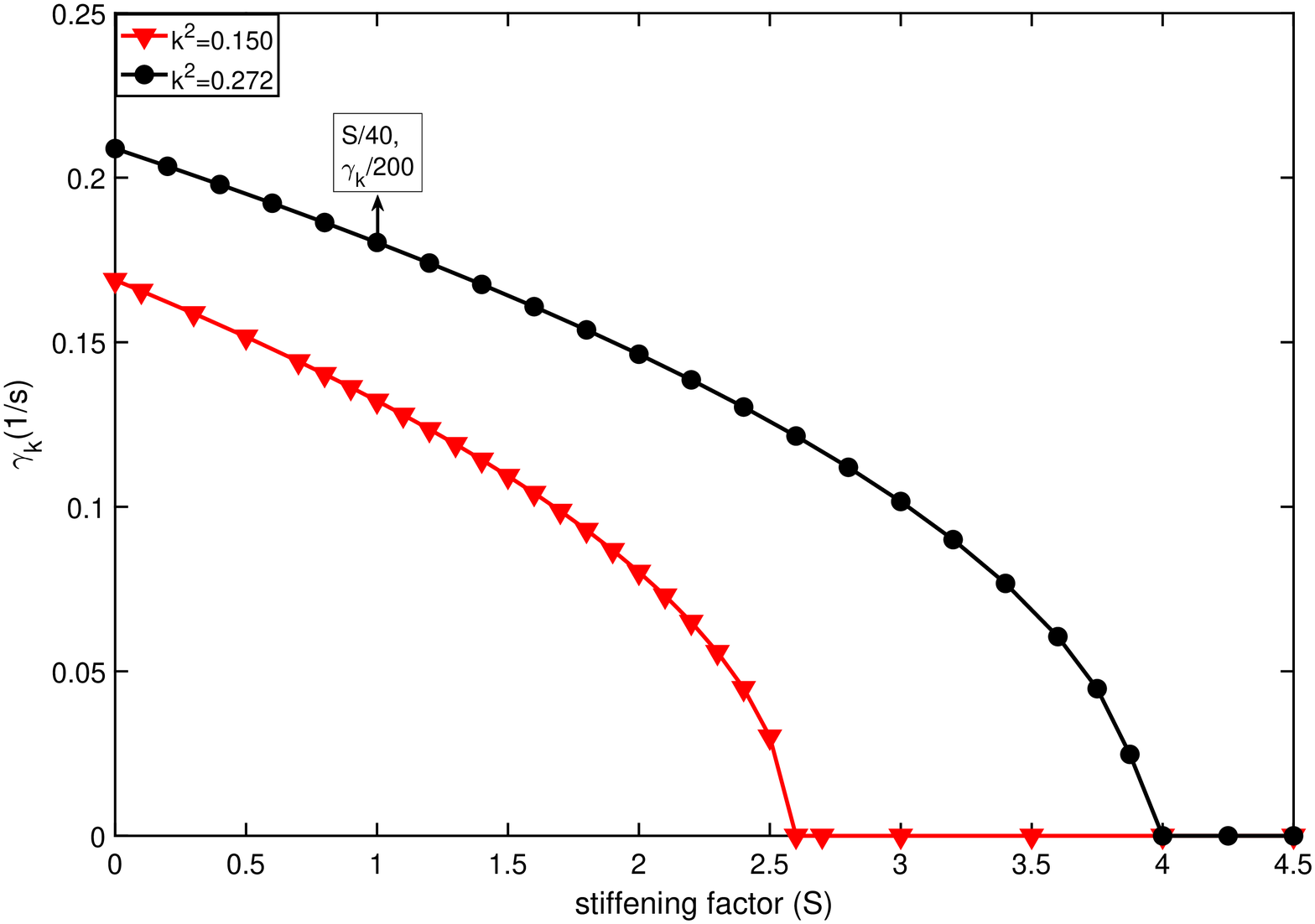}
 \caption{\label{pre_spit} The KBM growth rate as a function of stiffening factor $S$ from equation (3) for different equilibrium pressure scaling parameter $k^2$ values at $k_y=40R_E^{-1}$.}
\end{figure}
\clearpage
\begin{figure}[htbp]
\hspace*{-1.5cm}
\centering
\includegraphics[height=17 cm,width=19 cm]{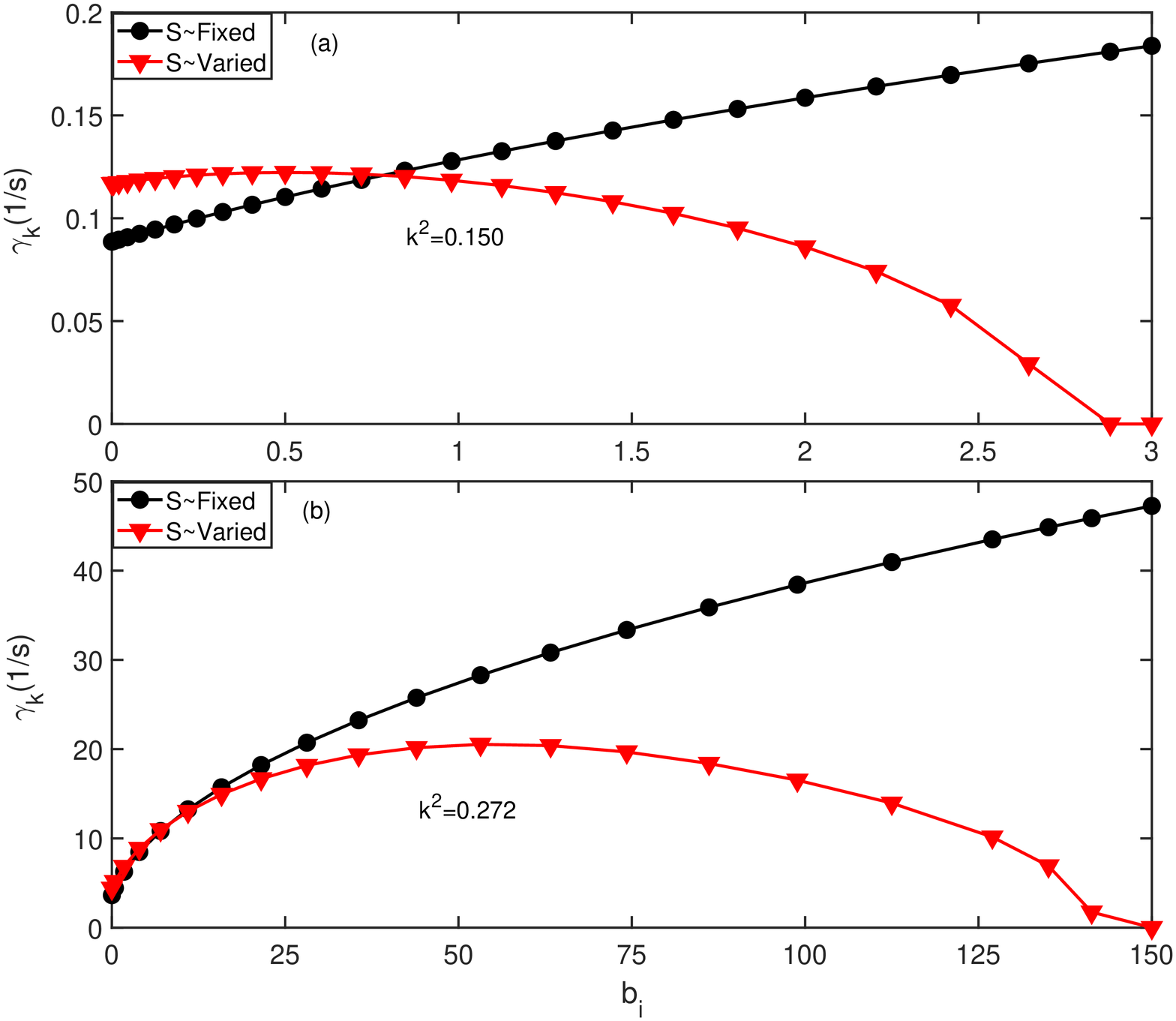}
\caption{\label{pre_spit}The KBM growth rate as a function of $b_i$ at the equatorial point $x_e=-9 R_E$ in the Voigt equilibrium model at $k_y=40 R_E^{-1}$ for (a) $k^2=0.15$ and (b) $k^2=0.272$.}
\end{figure}
\clearpage
\begin{figure}
\hspace*{-1.5cm}
\centering
\includegraphics[height=17 cm,width=19 cm]{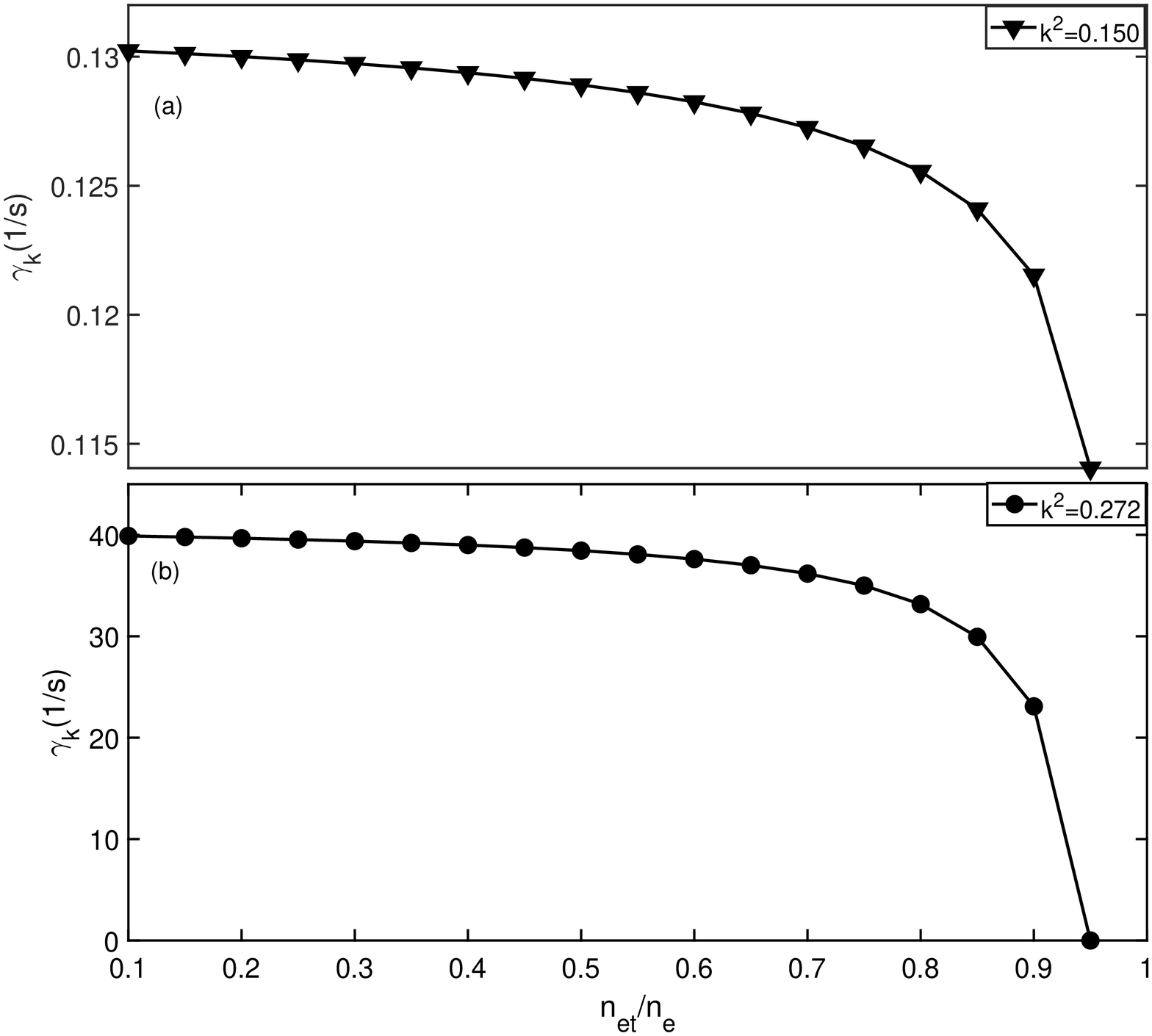}
 \caption{\label{pre_spit} The KBM growth rate as a function of $n_{et}/n_e$ in the Voigt equilibrium model at $k_y=40 R_E^{-1}$ for (a) $k^2=0.15$ and (b) $k^2=0.272$.}
\end{figure}
\clearpage
\begin{figure}[htbp]
\hspace*{-1.5cm}
\centering
\includegraphics[height=15 cm,width=19 cm]{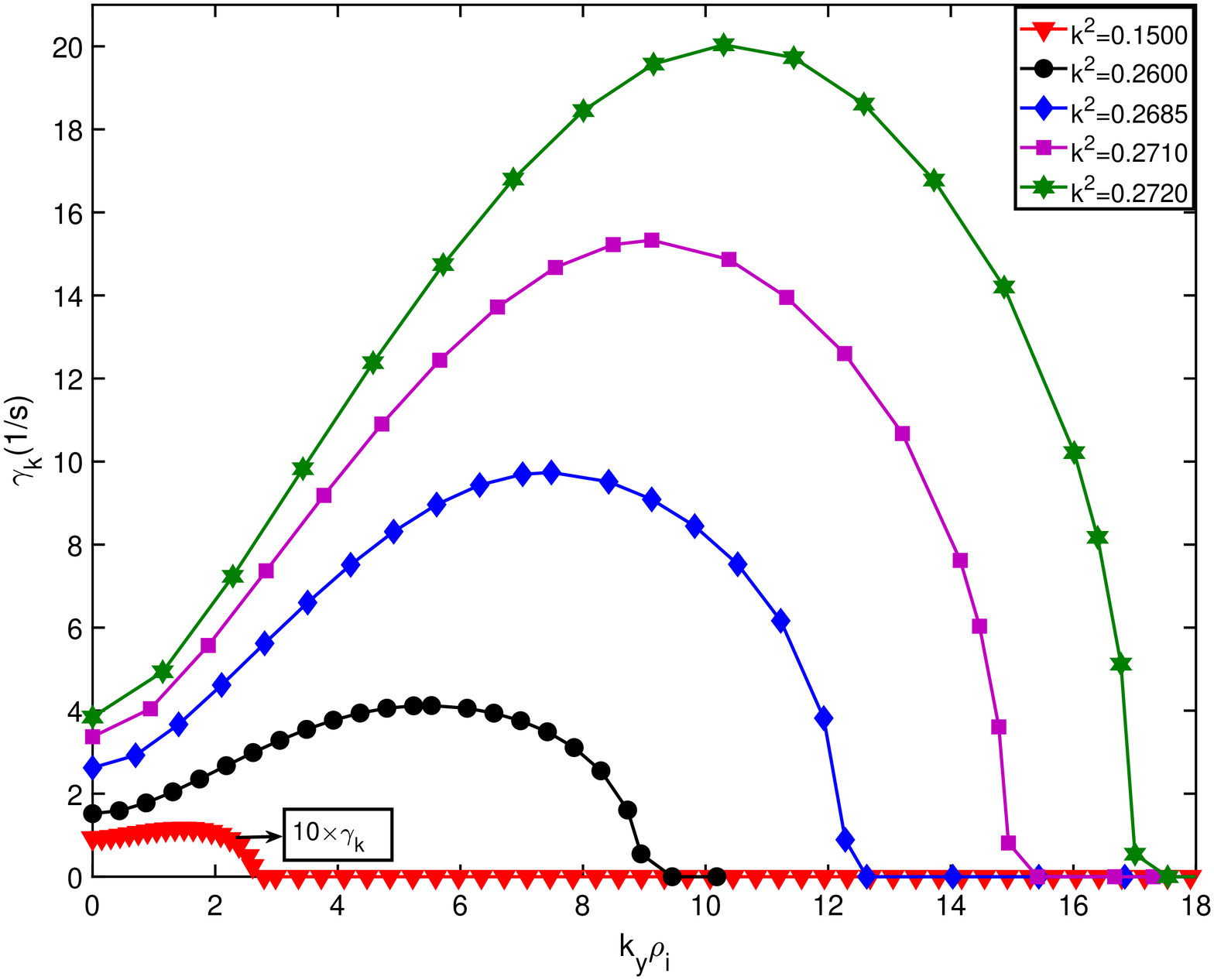}
\caption{\label{pre_spit} The KBM growth rate variation as a function of $k_y\rho_i$ at the equatorial point $x_e=-9 R_E$ in the Voigt equilibrium model with increasing values of pressure scaling parameter $(k^2)$.}
\end{figure}
\clearpage
\begin{figure}[htbp]
\hspace*{-1.5cm}
\centering
\includegraphics[height=17 cm,width=19 cm]{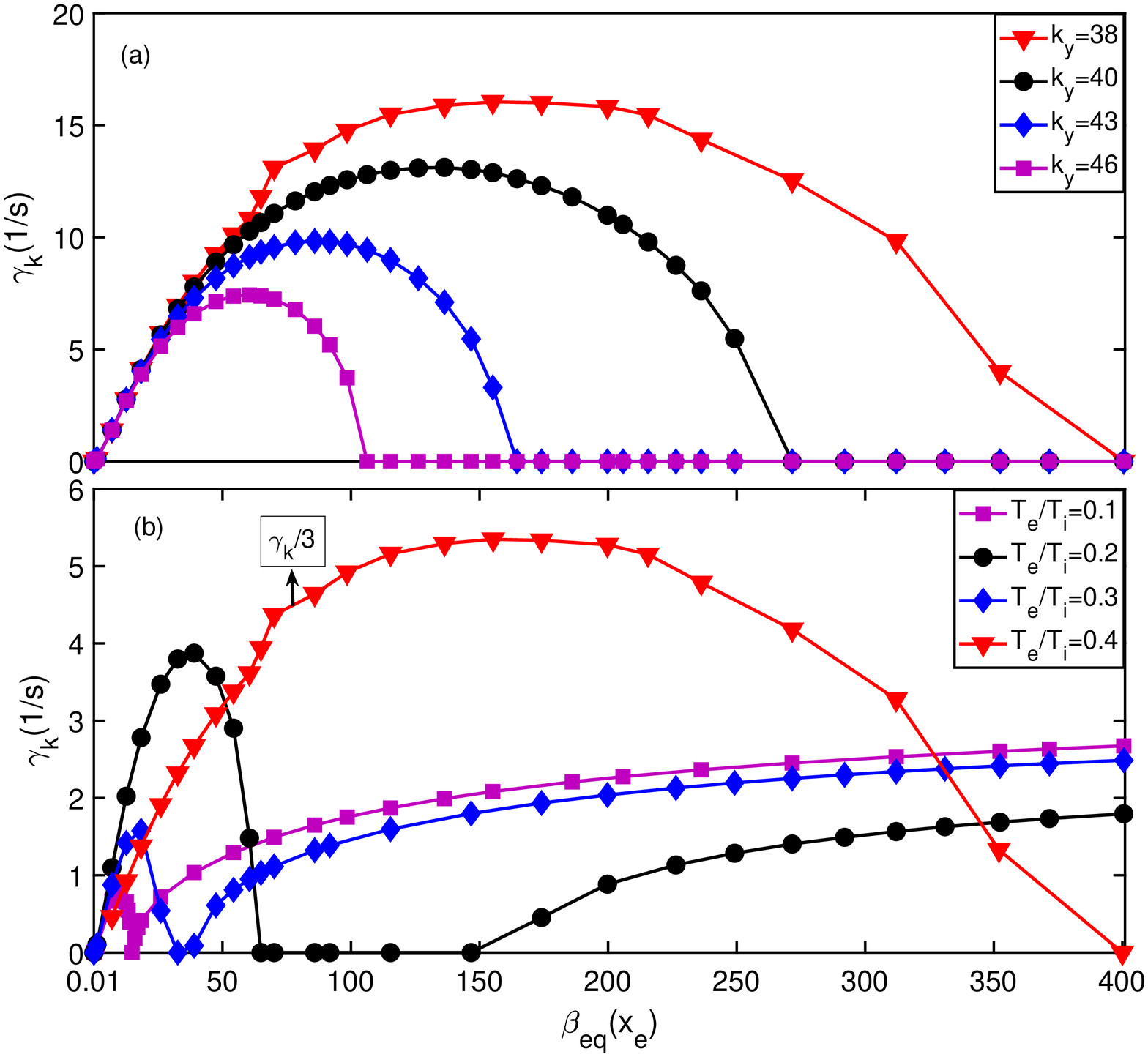}
\caption{\label{pre_spit}The KBM growth rate as a function of $\beta_{eq}$ for (a) four different values of $k_y(R_E^{-1})$ at $T_e/T_i=0.1$  and (b) four different values of $T_e/T_i$ at the equatorial point $x_e=-9 R_E$ for $k_y=38 R_E^{-1}$ in the Voigt equilibrium model.}
\end{figure}
\clearpage
\begin{figure}[htbp]
\hspace*{-1.5cm}
\centering
\includegraphics[height=19 cm,width=19 cm]{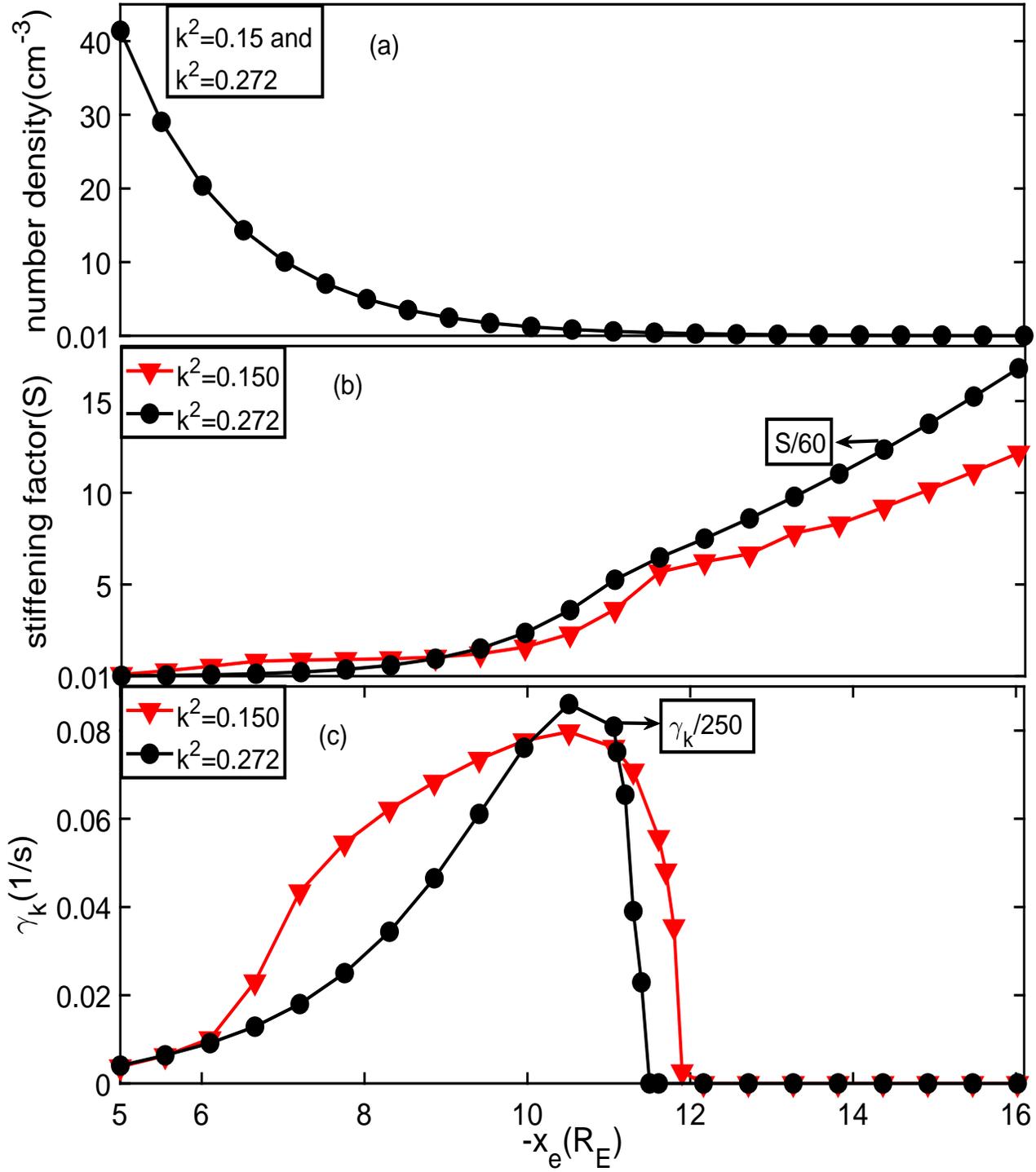}
\caption{\label{pre_spit} (a) Profile of the number density at $k^2=0.15$ with $T_i=1 keV$, (b) the stiffening factor $(S)$, and (c) maximum growth rate of the KBM as a function of $x_e$ at $k^2=0.15$ and $k^2=0.272$ using the Voigt equilibrium model with $k_y=38(R_E^{-1})$.}
\end{figure}
\clearpage
\begin{figure}[htbp]
\hspace*{-1.5cm}
\centering
\includegraphics[height=15 cm,width=19 cm]{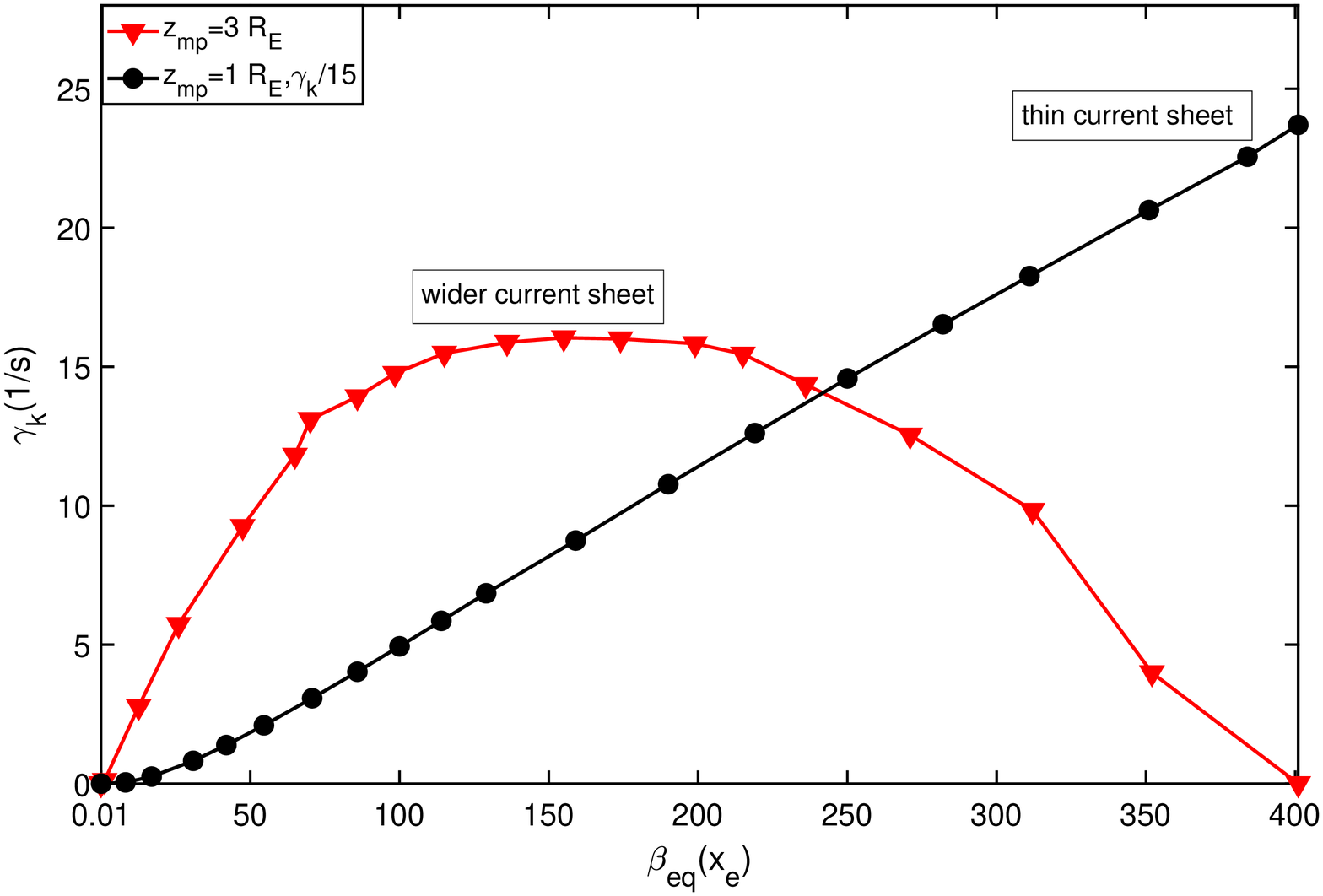}
\caption{\label{pre_spit} Comparison of the KBM growth rate between a thin $(z_{mp}=1 R_E)$ and a wider $(z_{mp}=3 R_E)$ current sheet configurations over a broad range of $\beta_{eq}$ at the equatorial point $x_e=-9 R_E$ in the Voigt equilibrium model for $k_y=38R_E^{-1}$.}
\end{figure}
\clearpage

\end{article}


%
%
%
%
%
%





\end{document}